\documentclass[
    ,final            
  ]
  {aipproc}

\layoutstyle{6x9}

\usepackage{amsmath,amssymb,bm}

\newcommand{\ra}{{\rightarrow}}
\newcommand{\vev}[1]{\langle #1 \rangle}
\newcommand{\Tr}{\mbox{Tr}}
\begin{document}

\title{Target mass corrections and beyond
\footnote{Talk given at ``SPIN2008'', University of Virginia,
October 6-11 2008.}
}

\classification{12.38.Bx, 13.60.Hb}
\keywords      {polarized DIS, target mass corrections, collinear factorization, OPE}

\author{Alberto Accardi}{
  address={Hampton University, Hampton, VA 23668, USA
  and Jefferson Lab, Newport News, VA 23606, USA} 
}

\begin{abstract}
I examine the uncertainty of perturbative QCD factorization for
(un)polarized hadron structure functions in deep inelastic scattering
at a large value of the Bjorken variable $x_B$. 
The focus will be on Target Mass Corrections and Jet Mass
Corrections in the collinear factorization framework.
\end{abstract}

\begin{flushright} JLAB-THY-08-930 \end{flushright}
\vskip-.5cm
\maketitle


Precise  parton distribution functions (PDFs) at large
parton fractional momentum $x$
are vital for understanding the nonperturbative
structure of the nucleon, and the effects of color confinement on its
partonic constituents.  For instance, the $d/u$ quark distribution ratio
near $x=1$ is very sensitive to the nature of the quark-quark forces in
the nucleon, as are parton-parton correlations in the nucleon wave
function.  Moreover, the magnitude and sign of the ratios of
spin-polarized to spin-averaged PDFs, $\Delta u/u$ and particularly
$\Delta d/d$, in the limit $x \to 1$ reflect the nonperturbative
quark-gluon dynamics in the nucleon, and can shed light on the
origin of the nucleon's spin.
Improved PDFs at large $x$ will be crucial at the upcoming 12~GeV
Jefferson Lab Upgrade for the planning and analysis of a host of
experiments, as well as at upcoming neutrino oscillation experiments
such as MINOS and MINER$\nu$A at Fermilab to control and reduce
systematic errors to a minimum. They will also directly impact experiments at
hadron colliders such as the Large Hadron Collider at CERN, by
facilitating accurate extraction of new physics signals as excess
compared to benchmark perturbative QCD (pQCD) calculations, and by
better constraining PDFs at lower $x$ and larger $Q^2$ via pQCD evolution.

Parton distribution functions can be extracted from experimental data
through global QCD fits, which combine data from many different processes
and observables, and analyze them by means of pQCD calculations.
Currently, the unpolarized parton distributions are well-determined
for $x \lesssim 0.5$ for quarks and $x \lesssim 0.3$ for gluons.
To better constrain them at large $x$ it is necessary to study hard
scattering processes near kinematic thresholds, such as Deep Inelastic
Scattering (DIS) at large Bjorken invariant $x_B$ and low 4-momentum transfer
squared $Q^2$, or Drell-Yan (DY) lepton pair production at large
Feynman variable $x_F$.
The large-$x$ and low-$Q^2$ region is even more important in the case of
parton helicity and transversity distributions, which measure the parton's
spin distribution and correlation to its orbital angular momentum, since
much of the presently available experimental data lie in this kinematic
domain.  

Theoretical improvements in calculating
leading-twist perturbative cross sections are needed in
order to confidently access large-$x$ / low-$Q^2$ PDFs. In this talk, I use
the collinear factorization framework to formulate and quantify 
corrections to DIS structure functions 
arising from target and jet mass corrections 
\cite{Accardi:2008ne,Accardi:2008pc}. Threshold resummation, higher twists,  
quark-hadron duality in the pre-asymptotic domain,
and nuclear effects such as binding, Fermi
motion and nucleon off-shellness are left for future efforts.

Target mass corrections (TMC) for spin-averaged nucleon structure functions
were first considered by Georgi \& Politzer within the Operator
Product Expansion (OPE) \cite{DGP,Blumlein:1998nv}. For spin-dependent
scattering they were evaluated within the same OPE formalism in
Refs.~\cite{Blumlein:1998nv,TMC-OPE}.
One of the limitations of the OPE formulation of TMCs is the so-called 
``threshold problem'', in which the target mass corrected structure 
functions remain nonzero at $x_B \geq 1$.
This arises from the failure to consistently incorporate the elastic
threshold in moments of structure functions at finite $Q^2$, resulting 
in nonuniformity of the $Q^2 \to \infty$ and $n \to \infty$
limits, where $n$ is the rank of the moment.
As a consequence, after performing an inverse Mellin transform on the
moments, the structure functions acquire incorrect support
at large $x_B$ \cite{Tung}.
Various prescriptions have been considered to tame the unphysical behavior
as $x_B \to 1$. However, these approaches are not unique and sometimes
introduce additional complications.

An alternative approach, which avoids the threshold ambiguities from
the outset, involves formulating TMCs directly in momentum space
using the collinear factorization (CF) formalism \cite{Accardi:2008ne}.
A further advantage of this formalism is that it can be
readily extended to processes such as semi-inclusive DIS, where the OPE
is not available, and indeed to any other hard scattering process.
Additional corrections to structure functions at large $x_B$, such as
discussed in the introduction, can also be
naturally incorporated together with TMCs. 
At any order in the strong coupling constant $\alpha_s$, and
considering only massless 
partons, the CF target-mass corrected transverse and
longitudinal structure functions, $F_{T}$ and $F_{L}$, read
\begin{align}
  F_{T,L}(x_B,Q^2,M^2)
    = \sum_f \int_\xi^{\xi/x_B} \frac{dx}{x}
    h_{f|T,L}\left(\frac{\xi}{x},Q^2\right)
    \varphi_f(x,Q^2) \ ,
 \label{eq:FTL_naive}
\end{align}
where $f$ is the parton flavor, $M$ the target's mass, 
$\xi$ the Nachtmann scaling variable,
\begin{align}
  \xi 
      = \frac{2 x_B}{1 + \sqrt{1+\gamma^2}}
  \qquad \gamma^2 = 4x_B^2 M^2/Q^2 \ ,
\end{align}
$h_f$ the perturbatively calculable coefficients, and 
$\varphi_f$ the unpolarized PDFs.
The polarized $g_1$ structure function reads 
\begin{align}
g_1(x_B,Q^2,M^2) &= \frac{1}{1+\gamma^2} \sum_f 
	    \int_\xi^{\xi/x_B} \!\frac{dx}{x}\,
	    g_{1,f}\left(\frac{\xi}{x},Q^2\right)\
	    \Delta\varphi_f(x,Q^2)\ ,
\label{eq:pTMC_CF1}
\end{align}
where $g_{1,f}$ are the perturbatively calculable coefficients, and 
$\Delta\varphi_f$ are the parton helicity distributions. The upper
limit of integration $x\leq\xi/x_B$ 
is imposed by combining four-momentum and baryon number
conservation in the kinematics of the handbag diagram. As a result,
the structure functions are zero when 
$x_B\geq1$, which solves the threshold problem. Note also that contrary
to previous claims  \cite{naiveCF}, 
$F_{T,L}(x_B) \neq F_{T,L}^{(0)}(\xi)$, and $g_1(x_B) \neq
(1+\gamma^2)^{-1}g_1^{(0)}(\xi)$, where the superscript $(0)$
indicates  ``massless'' structure functions computed with $M=0$, and
the upper limit on the $x$ integration is  $x\leq1$. 

\begin{figure}[t]
\centering
\parbox[c]{0.33\linewidth}{
\includegraphics
    [width=\linewidth,,trim=23 0 28 0,clip=true]
    {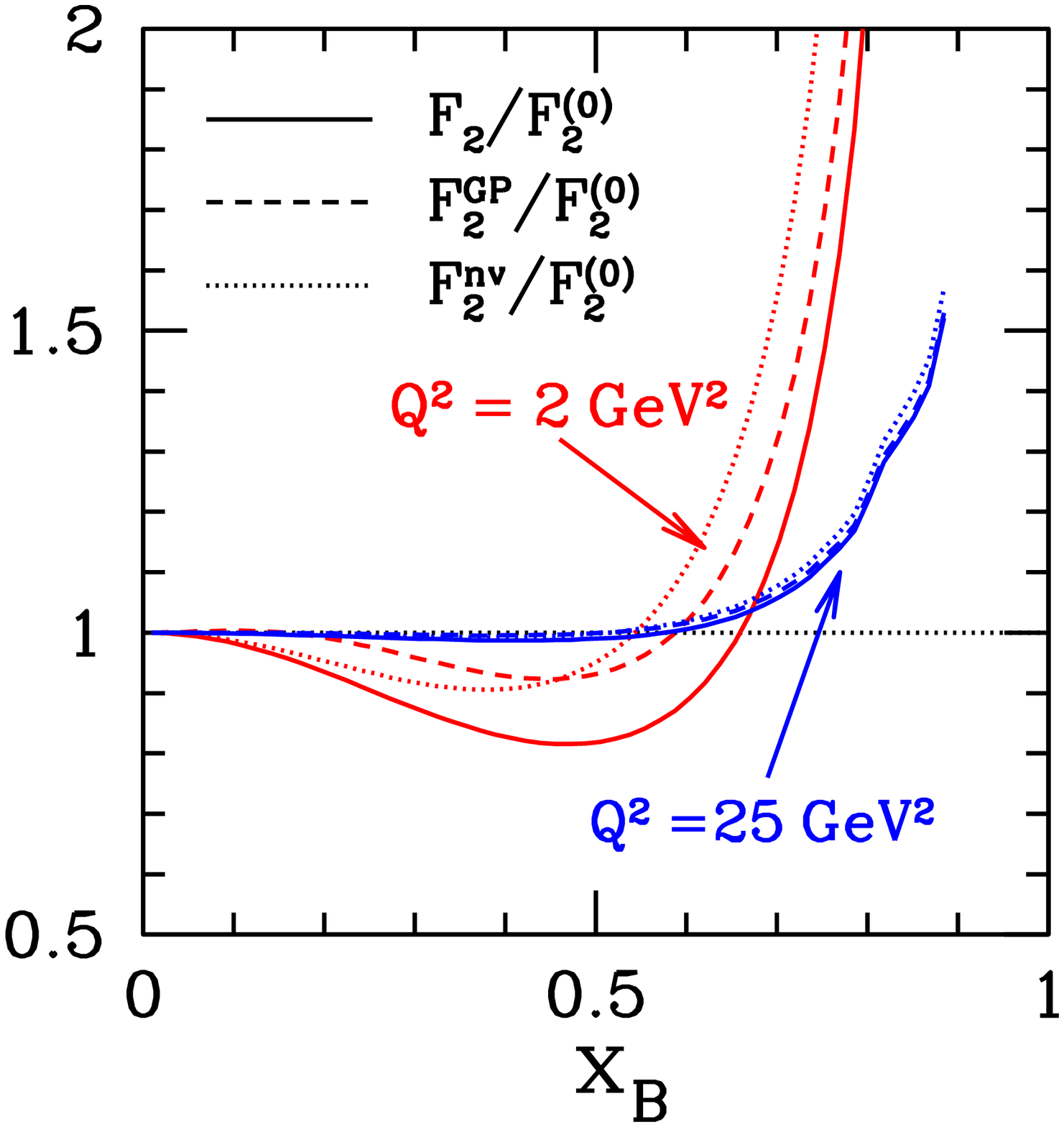}
\includegraphics
    [width=\linewidth,trim=23 0 28 0,clip=true]
    {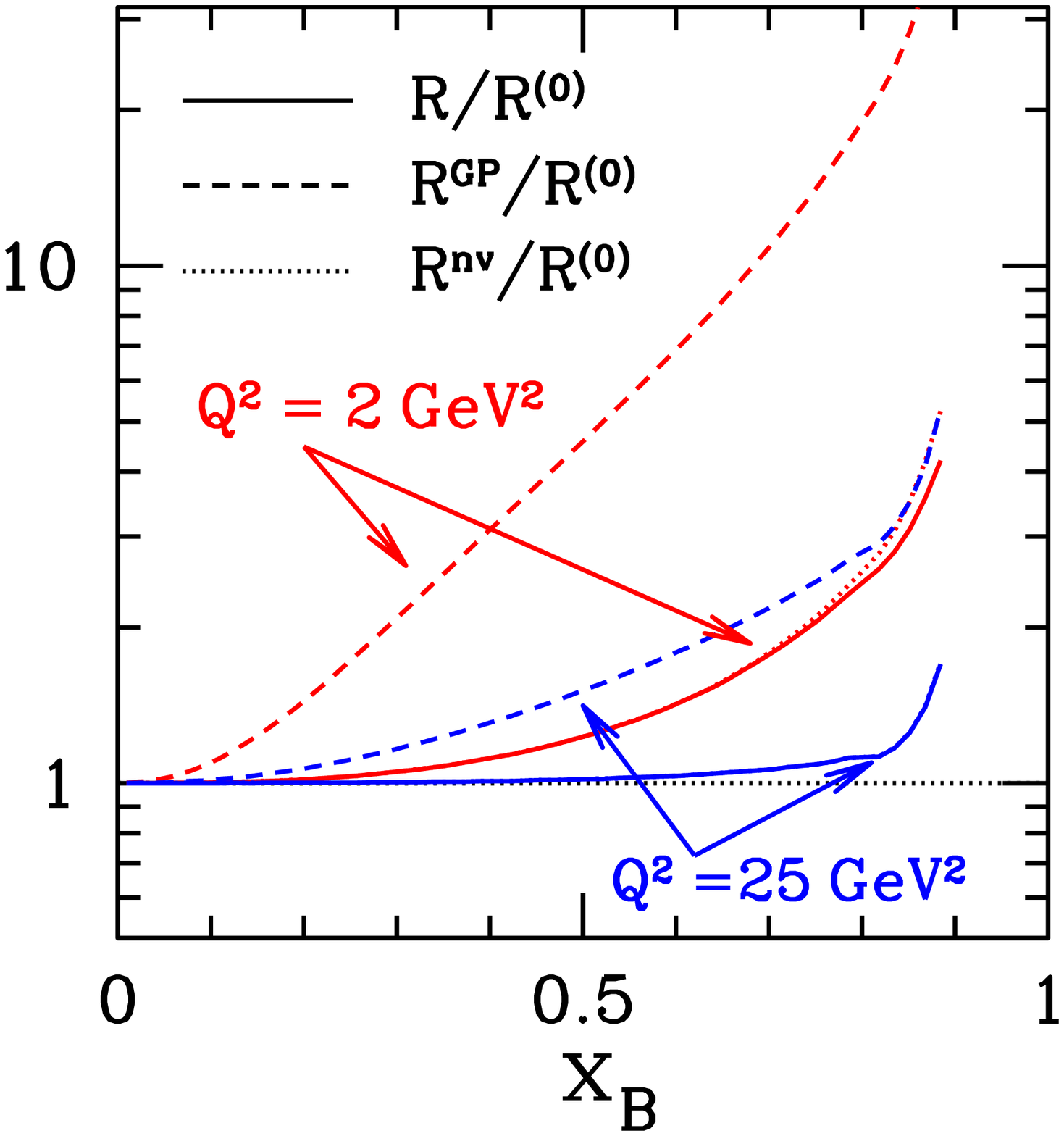}
}
\hspace*{0.3cm}
\parbox[c]{0.63\linewidth}{
\centering
\includegraphics
    [width=\linewidth,trim=0 15 20 -20,clip=true]
    {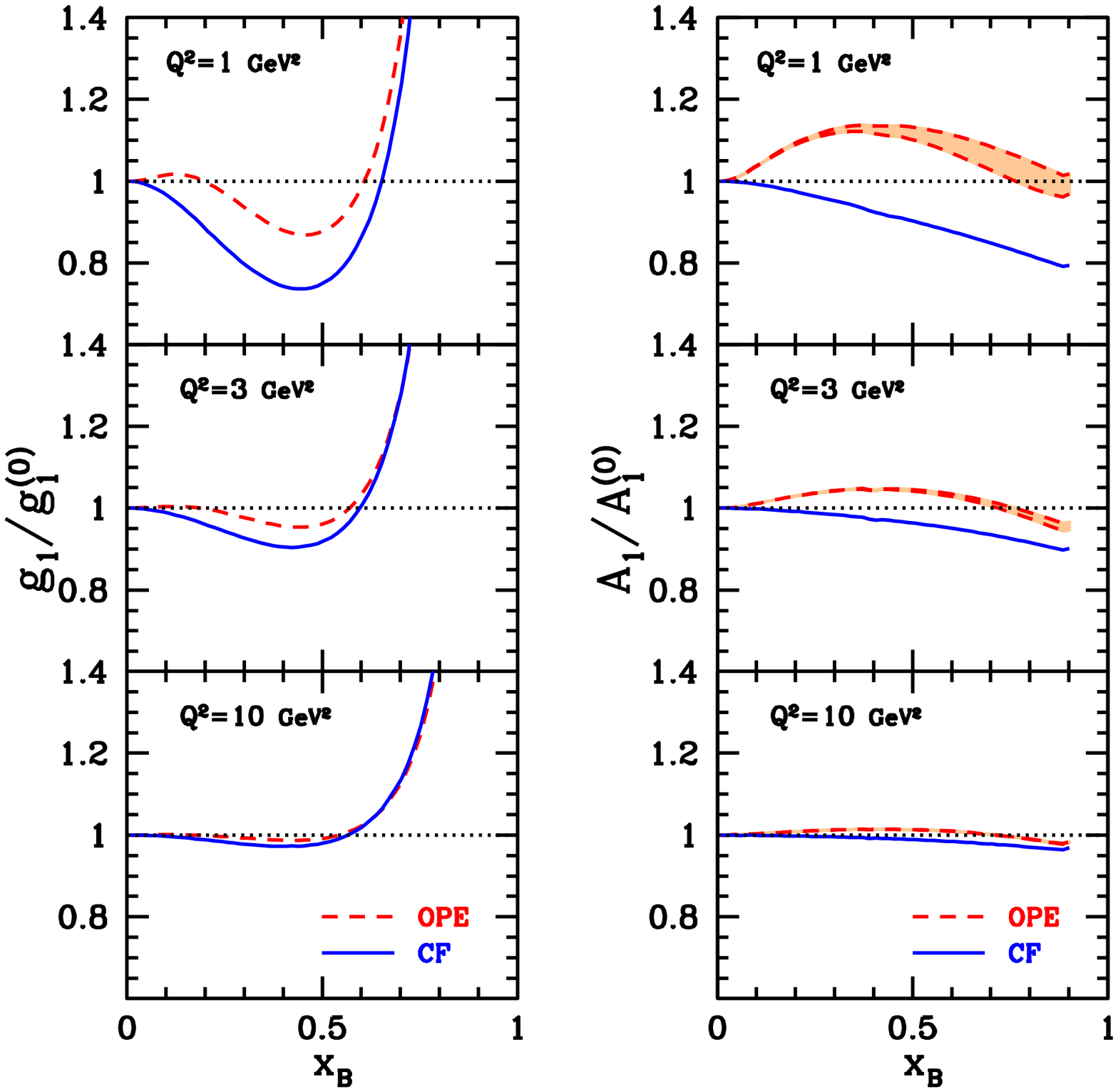}
  \caption{
    {\it Square plots:}
    Comparison of prescriptions for NLO target mass corrections to 
    $F_2$ and $R=F_L/F_1$, computed with MRST2002 parton
    distributions \cite{Accardi:2008ne}. 
    (The ``GP'' curves are computed in OPE; 
    the ``nv'' curves are computed using 1 as upper bound for the $x$
    integration in Eq.~\eqref{eq:FTL_naive}.
    {\it Vertical plots:} 
    Same for the LO $g_1$ structure function (left) and  $A_1$
    polarization asymmetry (right) \cite{Accardi:2008pc}. 
    For $A_1$, the shaded band for the OPE result indicates the
    uncertainty due to the use of the Wandzura-Wilczek relation 
    (lower bound) or the identity $g_1+g_2=0$ (upper bound) 
    in deriving $A_1$.
  }
  \label{fig:TMC}
}
\end{figure}

An estimate of TMC in collinear factorization for polarized and unpolarized
structure functions, and for the $A_1 = (1+\gamma^2) g_1/F_1$ asymmetry is
presented in Fig.~\ref{fig:TMC}, and compared to the TMC computed in OPE.
The numerical results for the target mass corrections to the $F_2$ and
$g_1$ structure
functions are qualitatively similar in the CF and OPE computations.
At leading order (LO), the differences of up to 30\% for low $Q^2$
values disappear for $Q^2 \gtrsim 10$ GeV$^2$. 
At next-to-leading order (NLO), the differences are generally larger
than at LO for intermediate $x_B$ values.   
In all cases the TMCs remain significant at $x_B \gtrsim 0.7$ even for
$Q^2 > 10$~GeV$^2$, and need to be taken into account when analyzing
large-$x_B$ data. In the longitudinal $F_L$ structure function
the CF and OPE corrections are larger by themselves, and 
differ by more than factors of several at large and not-so-large
$x_B$, even at large $Q^2$ values usually considered TMC-free. 
The choice of TMC scheme will affect, e.g., the determination of
the gluon PDF, to which $F_L$ is much more sensitive than $F_2$, and
is not a negligible matter. 
Since the TMCs are qualitatively similar for the $g_1$ and $F_1$ 
structure functions, they largely cancel in the $A_1$ asymmetry,
although the sign of the correction is opposite in the CF and OPE 
approaches over most of the range of $x_B$, but can be as large as 20\% at
low $Q^2$ , disappearing at $Q^2 \gtrsim 10$~GeV$^2$. 
The commonly used approximation relating $A_1$ directly to the
longitudinal asymmetry $A_\parallel$ can be shown to be equivalent to
assuming $A_1 \approx A_1^{(0)}$ \cite{Accardi:2008pc}. Fig.~\ref{fig:TMC} 
shows that it breaks down at
low $Q^2$: an accurate measurement of $g_1$, e.g., at JLab, 
requires both $A_\parallel$ and $A_\perp$
asymmetries.
 
Even after solving the threshold problem as described, the structure functions
remain unphysically positive as $x_B \ra 1$ because the assumption
that at LO the final state current jet is made of a single massless
quark makes the LO perturbative coefficients 
proportional to $\delta(x-\xi/x_B)$. Such an assumption is clearly
unphysical because the quark has to hadronize due to color confinement, 
so that the current jet will have an invariant mass $m_j^2$. Using
this mass in the kinematics of the LO handbag diagram, and assuming
that the invariant jet mass has a probability distribution $J_m$, 
one obtains 
\begin{align}
\begin{split}
 F_T^{JMC}(x_B,Q^2,M^2) & = \int_0^{\frac{1-x_B}{x_B}Q^2} 
   \hspace*{-.2cm}dm_j^2\, J_m(m_j^2) \,
   F_T^{(0)}\big(\xi(1+m_j^2/Q^2),Q^2\big) \ .
 \label{eq:FTL_TMC_JMC_heuristic}
\end{split}
\end{align}
If the ``jet function'' $J_m(m_j^2)$ is a sufficiently smooth function
of $m_j^2$, one obtains  $F_T^{JMC}(x_B,Q^2,M^2) \ra 0$ as $x_B\ra
1$, as it should be.
The jet mass corrections (JMC) so
introduced are of order $O(m_j^2/Q^2)$, and it is easy to see that in the
limit $Q^2 \gg \vev{m_j^2}$ one recovers the structure
functions with TMC only. It turns out that the ansatz
\eqref{eq:FTL_TMC_JMC_heuristic} is
theoretically correct, and that the jet function 
is the spectral function, $J_2$, of a vacuum quark propagator,
\begin{align}
  \int_0^\infty \hspace*{-.3cm}
    dm_j^2\,J_2(m_j^2)\,2\pi \delta(l^2-m_j^2)\, \theta(l^0)
    =  \frac{1}{4 l^-} \int d^4z e^{iz\cdot l}
    \Tr \big[ \gamma^- 
    \vev{ 0 | \overline\psi(z)\psi(0)|0} 
    \big] \ ,
\end{align}
additionally smeared by soft momentum exchanges with the target jet
\cite{Accardi:2008ne}.
A toy model estimate of these Jet Mass Corrections shows that they may
in fact be comparable in size to the TMC. However
their phenomenology is just at its beginning.


\begin{theacknowledgments}
This work was supported by the DOE contract No. DE-AC05-06OR23177,
under which Jefferson Science Associates, LLC operates Jefferson Lab,
and NSF award No. 0653508.
\end{theacknowledgments}




\begin{thebibliography}{9}


\bibitem{Accardi:2008ne}
  A.~Accardi and J.~W.~Qiu,
  JHEP {\bf 0807} (2008) 090.

\bibitem{Accardi:2008pc}
  A.~Accardi and W.~Melnitchouk,
  Phys.\ Lett.\  B {\bf 670} (2008) 114.

\bibitem{DGP}
  H.~Georgi and H.~D.~Politzer,
  Phys.\ Rev.\  D {\bf 14} (1976)1829;
%
  A.~De Rujula, H.~Georgi and H.~D.~Politzer,
  Annals Phys.\  {\bf 103} (1977) 315.

\bibitem{Blumlein:1998nv}
  J.~Blumlein and A.~Tkabladze,
  Nucl.\ Phys.\  B {\bf 553} (1999) 427.

\bibitem{TMC-OPE}
  S.~Matsuda and T.~Uematsu,
  Nucl.\ Phys.\  B {\bf 168} (1980) 181;
%
  A.~Piccione and G.~Ridolfi,
  Nucl.\ Phys.\  B {\bf 513} (1998) 301; 
%
  J.~Blumlein and A.~Tkabladze,
  Nucl.\ Phys.\  B {\bf 553} (1999) 427.

\bibitem{Tung}
  K.~Bitar, P.~W.~Johnson and W.~K.~Tung,
  Phys.\ Lett.\  B {\bf 83} (1979) 114;
  %
  P.~W.~Johnson and W.~K.~Tung,
  Print-79-1018 (Illinois Tech) {\it Contribution to Neutrino '79},
  Bergen, Norway, June 18-22, 1979.
%
  For a review, see 
  I.~Schienbein {\it et al.},
  J.\ Phys.\ G {\bf 35} (2008) 053101.

\bibitem{naiveCF}
  M.~A.~G.~Aivazis, F.~I.~Olness and W.~K.~Tung,
  Phys.\ Rev.\  D {\bf 50} (1994) 3085;
%
  S.~Kretzer and M.~H.~Reno,
  Phys.\ Rev.\  D {\bf 69} (2004) 034002. 

\end{thebibliography}
\end{document}